\documentclass[final,1p,times]{elsarticle} 
\usepackage{graphicx} 
\usepackage{epstopdf}
\usepackage{amssymb} 
\usepackage{amsthm} 
\usepackage{lineno} 
\newcommand{\sNN}[1]{$\sqrt{s_{NN}} = #1$ GeV}
\newcommand{\Npart}{$\langle N_{part}\rangle$}

\journal{Nuclear Physics A} 
\begin{document} 

\begin{frontmatter} 

\title{Strangeness Production in Heavy-Ion Collisions at STAR}

\author{Anthony R. Timmins$^{a}$ for the STAR Collaboration}

\address[a]{Department of Physics and Astronomy, Wayne State University, 666 W. Hancock, Detroit, MI 48201, USA}

\begin{abstract}
We report an overview of strangeness production in Cu+Cu and Au+Au  collisions at the energies $\sqrt{s_{NN}} =$ 62.4 and 200 GeV. We show new mid-rapidly $dN/dy$ results for the $K^{0}_{S}$, $\Lambda$, $\Xi$, $\Omega$ particles in Cu+Cu \sNN{62} collisions and compare to results in Au+Au \sNN{62} collisions. We show new results for mid-$p_T$ $\Lambda/K^{0}_{S}$ ratios in Cu+Cu \sNN{62} collisions and again compare to ratios in Au+Au \sNN{62} collisions. Finally, we show the high-$p_{T}$ ($\sim 6.2$ GeV/c) $R_{AA}(K^{0}_{S})$ as a function of system size in Au+Au \sNN{200} collisions and compare to $R_{AA}(\pi)$.
\end{abstract}

\end{frontmatter} 



\section{Introduction}

Measurements of strangeness production in heavy-ion collisions were originally conceived to be the smoking gun of Quark Gluon Plasma (QGP) formation \cite{RafalMull1, RafalMull2}. It was argued that due to a drop in the strange quark's dynamical mass and increased production cross section, strangeness in the QGP would saturate on small time scales relative to a hadronic gas. Such an abundance of strangeness quarks may lead to the bulk strange hadrons being enhanced in heavy-ion collisions relative to p+p, and allow for production mechanisms such as coalescence or jet flavor conversions to contribute to strange hadron yields in heavy-ion collisions at mid and high-$p_{T}$ respectively. 

In these proceedings, we show new results of mid-rapidity $dN/dy$ for the $K^{0}_{S}$, $\Lambda$, $\Xi$, $\Omega$ particles in Cu+Cu \sNN{62} collisions and compare to $dN/dy$ values in Au+Au \sNN{62} collisions. These yields are expected to be dominated by soft processes. 
We then move to new mid-$p_T$ measurements of the $\Lambda/K^{0}_{S}$ ratios in Cu+Cu \sNN{62} collisions where soft and hard processes may compete. Finally, we take advantage of new high-$p_T$ measurements of identified particle production in p+p \sNN{200} \cite{YXu}, and compare strange and non-strange Au+Au \sNN{200} $R_{AA}$ as a function of system size for $p_{T} > 5.5 $ GeV/c where hard processes are expected to dominate.

\section{Results}

The \emph{enhancement factor} for a strange hadron is defined as the yield per participant, \Npart, in heavy-ion collisions, divided by the respective value in p+p collisions at the same center of mass energy per nucleon. It contrasts strangeness production per unit of available energy (\Npart) in heavy-ion collisions with p+p collisions. When the enhancement factor is observed to be above one, this indicates the dynamical processes in heavy-ion collisions are much better at liberating the available energy for strangeness production compared to p+p collisions. Indeed, the enhancement factors for strange hadrons have been shown to be significantly above one for Pb+Pb \sNN{17.3}, Cu+Cu \sNN{200}, and Au+Au \sNN{200} collisions with a hierarchy with respect to the strangeness content \cite{StrangenessScaling, HelensPRC, HotQuarks}. Whether this is due to QGP formation as per the introduction, or phase space suppression in smaller systems \cite{CanoncialSuppression}, is an on going debate \cite{HotQuarks, Phi}.


\begin{figure}[h]
\begin{tabular}{c c}
\includegraphics[width = 0.5\textwidth]{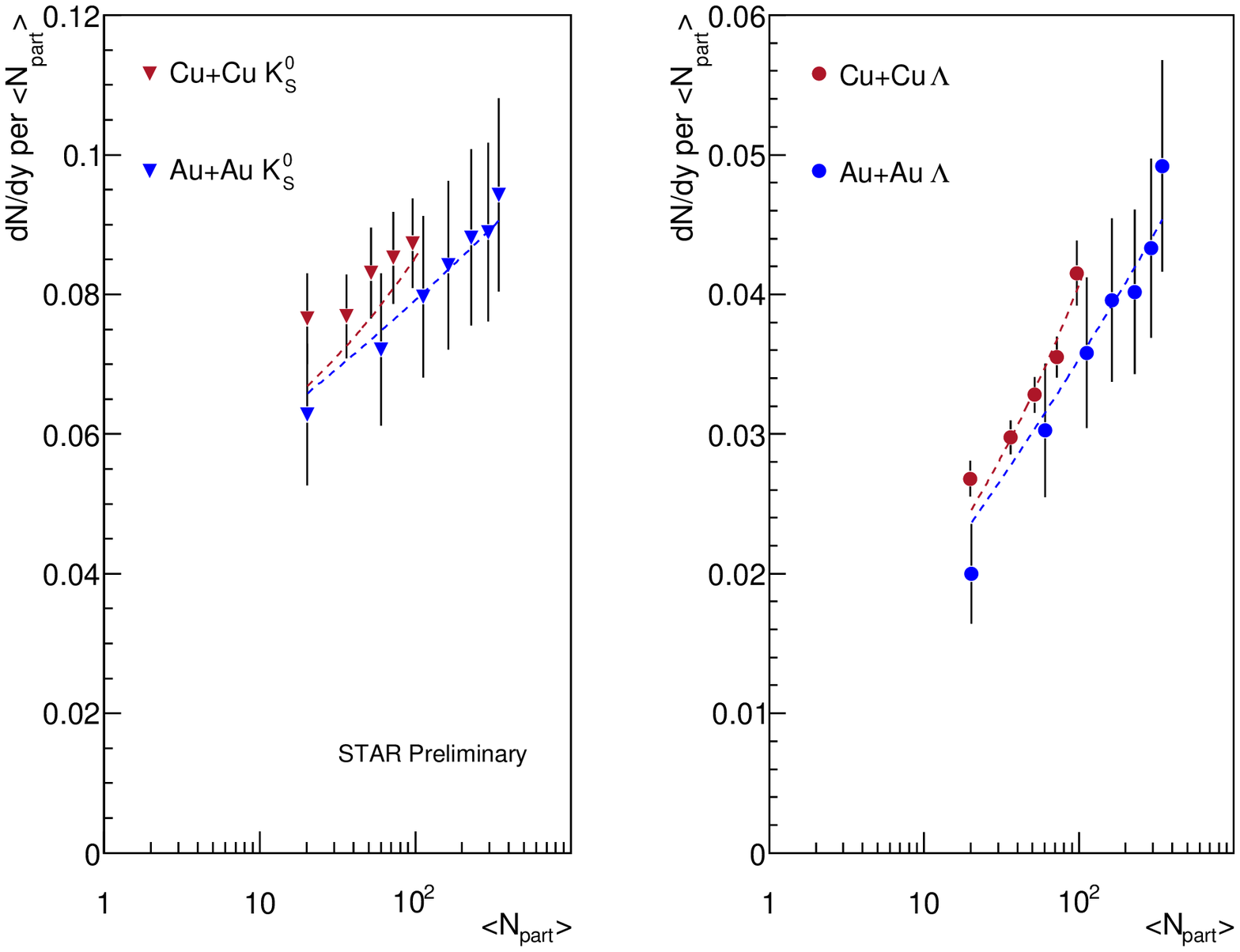}
&
\includegraphics[width = 0.5\textwidth]{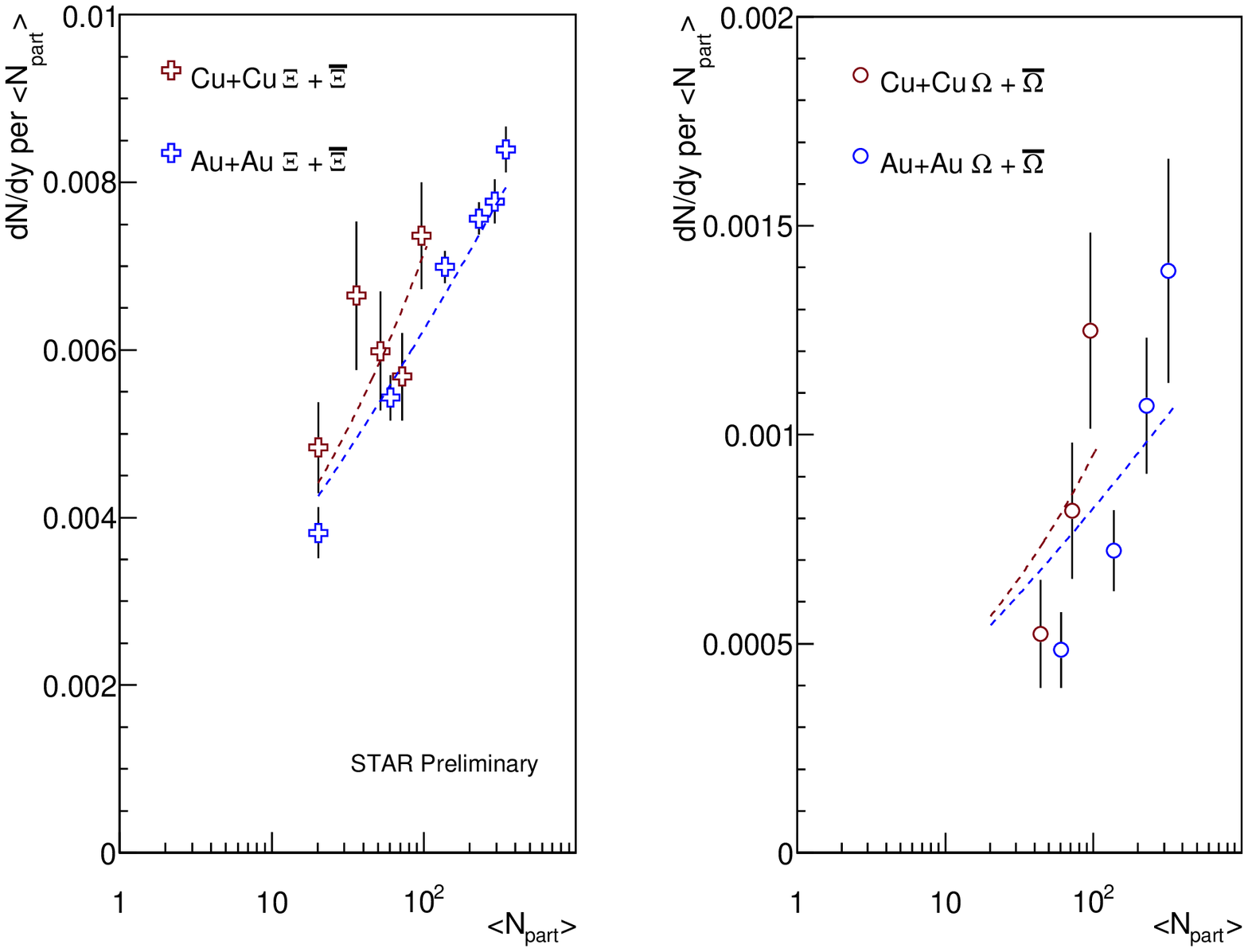}
\end{tabular}
\caption{$dN/dy$ per \Npart$ $ for the  $K^{0}_{S}$, $\Lambda$, $\Xi$, $\Omega$ particles in Cu+Cu and Au+Au \sNN{62} collisions. The uncertainties on the Cu+Cu data points are the statistical uncertainties from the yield and the systematic uncertainties from \Npart$ $ added in quadrature. The systematic uncertainties regarding the Cu+Cu yields are under investigation. The uncertainties on the Au+Au data are the full statistical and systematic uncertainties added in quadrature. The $\Lambda$ yields have not subtracted for feeddown from $\Xi$ decays; this contribution is typically $\sim13\%$. The curves are described in the text.}
\label{Fig1} 
\end{figure}

In figure \ref{Fig1}, we show $dN/dy$ per \Npart$ $ for the  $K^{0}_{S}$, $\Lambda$, $\Xi$, $\Omega$ particles in Cu+Cu and Au+Au \sNN{62} collisions. We are unable to calculate an enhancement factor due to the lack of a p+p reference. Becattini and Manninen have recently proposed $dN/dy$ per \Npart(therefore the enhancement factor) may be proportional to the fraction of participants that undergo multiple collisions, $f$, in a core-corona description of strangeness production for Au+Au \sNN{200} collisions \cite{phiCoreCorona1}. Hadron production from the core gives strangeness yields expected in the QGP saturation scenario, while corona production is p+p like. The curves in figure \ref{Fig1} are from the following relation:
\begin{equation}
\label{equ:fraction}
dN/dy \textrm{ \emph{per} } \langle N_{part} \rangle = D_{i}f(N_{part})+E_{i}
\end{equation}
where $D_{i}$ and $E_{i}$ are constants for particle $i$ chosen to describe the data. Those constants are independent of system for a given particle. $f(N_{part})$ is obtained from Monte Carlo Glauber calculations and is typically higher in Cu+Cu compared to Au+Au at a given \Npart. It is clear that the above relation provides a reasonable description of $dN/dy$ per \Npart$ $ in Cu+Cu and Au+Au \sNN{62} collisions. However, it was also shown that the above relation is not unique to the core-corona scheme, as it is expected for string excitation/breaking hadron production models \cite{HotQuarks}.

\begin{figure}[h]
\begin{center}
\includegraphics[width = 0.7\textwidth]{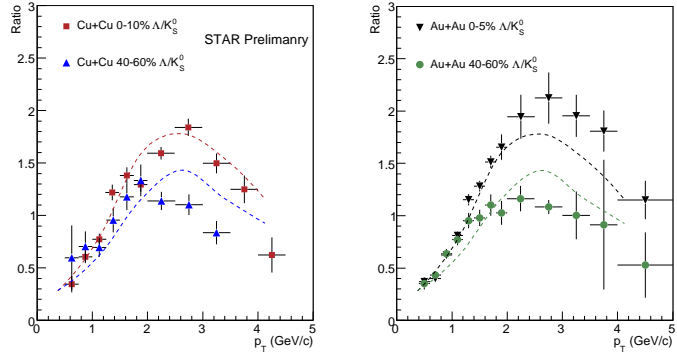}
\end{center}
\caption{The $\Lambda/K^{0}_{S}$ ratios as a function of $p_{T}$ for Cu+Cu and Au+Au \sNN{62} collisions. Uncertainties are statistical and the $\Lambda$ yields have not subtracted for feeddown from $\Xi$ decays. The curves are predictions from the EPOS model, version 1.67.}
\label{Fig2}
\end{figure}

In figure \ref{Fig2}, we show the $\Lambda/K^{0}_{S}$ ratios as a function of $p_{T}$ for Cu+Cu and Au+Au \sNN{62} collisions. Large values at mid-$p_{T}$ in central Au+Au \sNN{200} collisions maybe attributed to quark coalescence boosting baryon production relation to meson production \cite{AuAuK0S, BarMes1, BarMes2, BarMes3}. At the lower energy, we observe central values at mid-$p_{T}$ being roughly a factor of 2 higher than peripheral values. 
To widen current theoretical comparisons, we show predictions from the EPOS model which also describes particle production in the core-corona scheme \cite{EPOS1}. After formation, the core is given a blast-wave velocity profile which boosts the $p_{T}$ of heavier particles relative to the lighter particles upon hadronization. Corona production is again p+p like, and the relative core contribution to particle production increases with centrality. 
We can see EPOS gives a qualitative description of the data. This was also shown for the $\Lambda/K^{0}_{S}$ ratios  in Cu+Cu and Au+Au \sNN{200} collisions. In particular, it predicts the mid-$p_{T}$ $\Lambda/K^{0}_{S}$ ratio should be approximately the same in central Cu+Cu and Au+Au collisions despite the large difference in \Npart ($\sim96$ vs. $\sim 344$ respectively). Finally, it was shown EPOS described the $\Omega/\phi$ ratios \cite{OmegaPhi1, OmegaPhi2, OmegaPhi3} well in Cu+Cu and Au+Au collisions with \sNN{200}. 

\begin{figure}[h]
\begin{center}
\includegraphics[width = 0.5\textwidth]{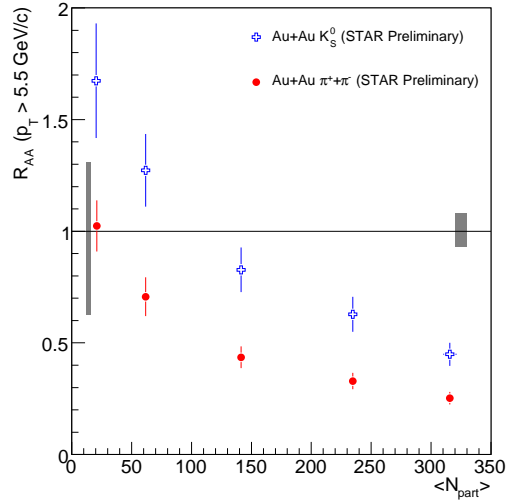}
\end{center}
\caption{Integrated $R_{AA}(K^{0}_{S})$ and $R_{AA}(\pi)$ for $p_{T} > 5.5$ GeV/c) in Au+Au \sNN{200} collisions. The uncertainties on the data points are statistical and systematic added in quadrature with respect to the yields. The left and right grey bands represent the typical uncertainties on $\langle N_{bin} \rangle $ for peripheral and central Au+Au \sNN{200} collisions respectively.}
\label{Fig3}
\end{figure}

In figure \ref{Fig3}, we show the integrated $R_{AA}(K^{0}_{S})$ and $R_{AA}(\pi)$ in Au+Au \sNN{200} collisions for particles with $p_{T} > 5.5$ GeV/c. It this therefore defined as:

\begin{equation}
\label{equ:fraction}
R_{AA} (p_{T} > 5.5 GeV/c) = \frac{ dN_{A+A}/dy \textrm{ \emph{per} }  \langle N_{bin} \rangle }{ dN_{p+p}/dy}
\end{equation}

The mean $p_{T}$ of particles in this range is $\sim 6.2$ GeV/c. It has recently been proposed that hard scattered partons may change flavor in the medium altering jet chemistry compared to p+p collisions \cite{JetConver}. The abundance of strange quarks created in heavy-ion collisions allow for the leading parton to be replaced a strange quark via elastic interactions, and such a mechanism is predicted to boost $R_{AA}(K^{0}_{S})$ for particles with $p_{T} > 5 $ GeV/c in central Au+Au \sNN{200} collisions by a factor 2 relative to $R_{AA}(\pi)$. We indeed observe such an increase in figure \ref{Fig3}, however we also note that $R_{AA}(K^{0}_{S}) > R_{AA}(\pi)$ in peripheral collisions. This raises the question whether parton flavor conversions are prevalent in the smaller systems, or whether there is some other soft A+A production mechanism contributing in this $p_T$ range for all centralities. Measurements at higher $p_{T}$ from future higher statistics RHIC A+A runs may help in addressing these questions.

\section{Conclusions}

We have presented new results for the $K^{0}_{S}$, $\Lambda$, $\Xi$, $\Omega$ particles in Cu+Cu \sNN{62} collisions, and shown in conjunction with Au+Au 62  \sNN{62} data, the centrality dependance of $dN/dy$ per \Npart$ $ has similar trends to a parameterisation based on the fraction of participants that undergo multiple collisions. We also have shown the $p_{T}$ dependance of the $\Lambda/K^{0}_{S}$ ratio can be qualitatively described by the EPOS model in Cu+Cu and Au+Au collisions with $\sqrt{s_{NN}}=62$ and 200 GeV. Both observations provide evidence for core-corona effects. Finally, we have shown at $p_{T} > 5.5$ GeV/c $R_{AA}(K^{0}_{S}) > R_{AA}(\pi)$ for all centralities in Au+Au \sNN{200} collisions. This may provide evidence for parton flavor conversions in heavy-ion collisions.


\begin{thebibliography}{10}
\bibitem{RafalMull1} J Rafelski and B M\"{u}ller, Phys. Rev. Lett. 48 (1982) 1066 
\bibitem{RafalMull2} P Koch, B M\"{u}ller and J Rafelski, Phys. Rep. 142 (1986) 167
\bibitem{YXu} Y Xu for the STAR Collaboration, \emph{these proceedings}
\bibitem {StrangenessScaling} J Adams et al. (STAR Collaboration), Phys. Rev. Lett. 98 (2007) 62301
\bibitem {HelensPRC} B I Abelev et al. (STAR Collaboration), Phys. Rev. C 77 (2008) 044908   
\bibitem{CanoncialSuppression} S Hamieh, K Redlich and A Tounsi, Phys. Lett. B. 486 (2000) 61
\bibitem{HotQuarks} A Timmins for the STAR collaboration, Eur. Phys. J. C. 62 (2009) 249
\bibitem{Phi} B I Abelev et al. (STAR Collaboration), Phys. Lett. B 673 (2009) 183
\bibitem{phiCoreCorona1} F Becattini and J Manninen, J. Phys. G 35 (2008) 104013
\bibitem{AuAuK0S} J Adams et al. (STAR Collaboration), \emph{nucl-ex/0601042}
\bibitem{BarMes1} B I Abelev et al. (STAR Collaboration), Phys. Rev. Lett. 97 (2006) 152301
\bibitem{BarMes2} B I Abelev et al. (STAR Collaboration),  Phys. Lett. B 655, 104 (2007) 
\bibitem{BarMes3} J Adams et al. (STAR Collaboration), Phys. Rev. Lett. 92 (2004) 52302
\bibitem{EPOS1} K Werner, Phys. Rev. Lett. 98 (2007) 152301
\bibitem{OmegaPhi1} B I Abelev et al. (STAR Collaboration), Phys. Rev. Lett. 99 (2007) 112301 
\bibitem{OmegaPhi2} B I Abelev et al. (STAR Collaboration), Phys. Lett. B 655 (2007) 104
\bibitem{OmegaPhi3} X Wang for the STAR Collaboration, J. Phys. G 35 (2008) 104074
\bibitem{JetConver} W Liuand and R L Fries, Phys. Rev. C 77 (2008) 054902

\end{thebibliography}
\end{document}